\documentclass[twocolumn,a4paper]{revtex4}
\usepackage{amsfonts,amsmath,amssymb,mathrsfs,dsfont,yfonts,bbm}
\usepackage[colorlinks,allcolors=blue]{hyperref}
\usepackage{graphicx}
\begin{document}

\newcommand{\Supertwistor}{\Cset \mathrm{P}^{3|4}}
\newcommand{\Twistorspace}{\Cset \mathrm{P}^{3}}
\newcommand{\half}{\frac{1}{2}}
\newcommand{\diff}{\mathrm{d}}
\newcommand{\ra}{\rightarrow}
\newcommand{\Zset}{{\mathbb Z}}
\newcommand{\Cset}{{\,\,{{{^{_{\pmb{\mid}}}}\kern-.47em{\mathrm C}}}}}
\newcommand{\Rset}{{\mathrm{I}\!\mathrm{R}}}
\newcommand{\gra}{\alpha}
\newcommand{\grl}{\lambda}
\newcommand{\gre}{\epsilon}
\newcommand{\zb}{{\bar{z}}}
\newcommand{\mn}{{\mu\nu}}
\newcommand{\Acal}{{\mathcal A}}
\newcommand{\Rcal}{{\mathcal R}}
\newcommand{\Dcal}{{\mathcal D}}
\newcommand{\Mcal}{{\mathcal M}}
\newcommand{\Ncal}{{\mathcal N}}
\newcommand{\Kcal}{{\mathcal K}}
\newcommand{\Lcal}{{\mathcal L}}
\newcommand{\Scal}{{\mathcal S}}
\newcommand{\Wcal}{{\mathcal W}}
\newcommand{\Bcal}{\mathcal{B}}
\newcommand{\Ccal}{\mathcal{C}}
\newcommand{\Vcal}{\mathcal{V}}
\newcommand{\Ocal}{\mathcal{O}}
\newcommand{\Zcal}{\mathcal{Z}}
\newcommand{\Zb}{\overline{Z}}
\newcommand{\Urm}{{\mathrm U}}
\newcommand{\Srm}{{\mathrm S}}
\newcommand{\SO}{\mathrm{SO}}
\newcommand{\Sp}{\mathrm{Sp}}
\newcommand{\SU}{\mathrm{SU}}
\newcommand{\U}{\mathrm{U}}
\newcommand{\be}{\begin{equation}}
\newcommand{\ee}{\end{equation}}
\newcommand{\Comment}[1]{{}}
\newcommand{\tQ}{\tilde{Q}}
\newcommand{\tq}{{\tilde{q}}}
\newcommand{\trho}{\tilde{\rho}}
\newcommand{\tphi}{\tilde{\phi}}
\newcommand{\Qcal}{\mathcal{Q}}
\newcommand{\tmu}{\tilde{\mu}}
\newcommand{\dbar}{\bar{\partial}}
\newcommand{\p}{\partial}
\newcommand{\eg}{{\it e.g.\;}}
\newcommand{\ie}{{\it i.e.\;}}
\newcommand{\Tr}{\mathrm{Tr}}
\newcommand{\twistor}{\Cset \mathrm{P}^{3}}
\newcommand{\note}[2]{{\footnotesize [{\sc #1}}---{\footnotesize   #2]}}
\newcommand{\CL}{\mathcal{L}}
\newcommand{\CJ}{\mathcal{J}}
\newcommand{\CA}{\mathcal{A}}
\newcommand{\CH}{\mathcal{H}}
\newcommand{\CD}{\mathcal{D}}
\newcommand{\CE}{\mathcal{E}}
\newcommand{\CQ}{\mathcal{Q}}
\newcommand{\CB}{\mathcal{B}}
\newcommand{\CC}{\mathcal{C}}
\newcommand{\CO}{\mathcal{O}}
\newcommand{\CT}{\mathcal{T}}
\newcommand{\CI}{\mathcal{I}}
\newcommand{\CN}{\mathcal{N}}
\newcommand{\CS}{\mathcal{S}}
\newcommand{\CM}{\mathcal{M}}

 \newcommand{\jhl}[1]{ {\color{red} #1}}

\parskip 11pt
\title{\Large {\bf An Exact Operator Map from Strong Coupling to Free Fields: \\ Beyond Seiberg-Witten Theory}}
\author {Chinmaya Bhargava, Matthew Buican, and Hongliang Jiang} 
\affiliation{CTP and Department of Physics and Astronomy \\ Queen Mary University of London, London E1 4NS, UK\\ }

\begin{abstract}
In quantum field theory (QFT) above two spacetime dimensions, one is usually only able to construct exact operator maps from the ultraviolet (UV) to the infrared (IR) of strongly coupled renormalization group (RG) flows for the most symmetry-protected observables. Famous examples include maps of chiral rings in 4d $\CN=2$ supersymmetry. In this letter, we construct the first non-perturbative UV/IR map for less protected operators: starting from a particularly \lq\lq simple" UV strongly coupled non-Lagrangian 4d $\CN=2$ QFT, we show that a universal non-chiral quarter-BPS ring can be mapped exactly and bijectively to the IR. In particular, strongly coupled UV dynamics governing infinitely many null states manifest in the IR via Fermi statistics of free gauginos. Using the concept of arc space, this bijection allows us to compute the exact UV Macdonald index in the IR.
\end{abstract}
\maketitle

\section*{Introduction}
In order to gain insight into strongly coupled QFT, it is useful to construct universal and calculable observables. However, there is often tension: the less calculable an observable is, the more interesting the dynamics it can probe.

In the case of 4d $\CN=2$ QFTs, the half-BPS chiral ring is a calculable space of operators maximally protected by supersymmetry. Through the celebrated machinery of Seiberg-Witten (SW) theory \cite{Seiberg:1994aj,Seiberg:1994rs}, it can be followed exactly along strongly coupled RG flows to the IR, where it gives the two-derivative effective theory on a moduli space of vacua called the \lq\lq Coulomb branch." 

One longstanding open question in strongly coupled QFT in $d>2$ is to give an exact UV/IR map of non-chiral observables less protected by supersymmetry. In this letter, we solve this problem for a ring arising from normal-ordered products of superpartners of the energy-momentum tensor. Unlike the SW ring, this ring is non-chiral, quarter-BPS, and hence \lq\lq half" as protected by supersymmetry. Geometrically, these results give an infinite-dimensional generalized tangent space of the Coulomb branch.

Our approach is to first focus on the closest and simplest strongly coupled 4d analog of an exactly solvable 2d QFT: the original or \lq\lq minimal" Argyres-Douglas (MAD) superconformal field theory (SCFT) \cite{Argyres:1995jj}. Indeed, from the point of view of the Coulomb branch effective theory, this SCFT is maximally simple. It also has the simplest symmetry structure of any 4d $\CN=2$ SCFT.  Finally, parts of the local operator algebra are maximally simple for a unitary theory with a vacuum moduli space \cite{Liendo:2015ofa,Buican:2021elx,Bhargava:2022cuf,Bhargava:2022yik}\footnote{The $S^1$ reduction is a free twisted hypermultiplet \cite{Benvenuti:2018bav} and also maximally simple.}.

This \lq\lq closeness" of the MAD theory to the Coulomb branch effective theory and certain exact spectroscopic results \cite{Bhargava:2022cuf} prompted us to conjecture the local operator algebra is generated as follows \cite{Bhargava:2022cuf}
\begin{equation}\label{CConj}
\CO\in\bar\CE_{6/5}^{\times m}\times\CE_{-6/5}^{\times n}~,\ \ \ \forall\ \CO\in\CH_L~.
\end{equation}
Here $\CO$ is any local operator of the SCFT ($\CH_L$ is the corresponding Hilbert space), and the righthand side of the inclusion represents the $(m,n)$-fold operator product expansion (OPE) of $\bar\CE_{6/5}$ and $\CE_{-6/5}$. In the language of \cite{Dolan:2002zh}, $\bar\CE_{6/5}$ is the multiplet housing the dimension $6/5$ chiral primary whose vev parameterizes the Coulomb branch ($\CE_{-6/5}$ houses the conjugate anti-chiral primary). Turning on a vev for the corresponding primary initiates an RG flow to the Coulomb branch and, in the deep IR, to free super-Maxwell theory.  Since the multiplets generating the MAD operator algebra are, in this sense, \lq\lq Coulombic," we refer to the above conjecture as the \lq\lq Coulombic generation" of the spectrum.

Given \eqref{CConj}, it is natural to try relating all non-decoupling parts of the MAD spectrum to super-Maxwell operators. A first step is to consider the generating multiplets \eqref{CConj}. As described above, the RG map in this case follows from the SW construction \cite{Argyres:1995jj}
\begin{equation}\label{EDmap}
\bar\CE_{6/5}\longrightarrow\bar\CD^{\rm Free}_{0(0,0)}~,
\end{equation}
where the righthand side is the free vector multiplet housing the chiral $\phi$ primary \footnote{The primary, $\CO_{6/5}\in\bar{\CE}_{6/5}$ generates an infinite chiral ring ($\CO_{6/5}^n:=\CO_{6n/5}\in\bar{\CE}_{6n/5}$). Therefore, when substituting a vev into $\CO_{6n/5}$, we see that $\bar{\CE}_{6n/5}$ also flows to $\bar{\CD}^{\rm Free}_{0(0,0)}$ at leading order. In order to have a meaningful map, we subtract lower-dimensional operators that can mix with the multiplet in question on flows to the Coulomb branch. For example, when mapping $\bar{\CE}_{12/5}$ to the IR, we implicitly subtract a term proportional to $\bar{\CE}_{6/5}$ so that this multiplet maps to $\bar{\CE}^{\rm Free}_2$. \label{EDfoot}}.

Another natural representation to consider is the stress tensor multiplet, which appears in the $m=n=1$ OPE in \eqref{CConj}. Since the RG flow preserves $\CN=2$, we have
\begin{equation}\label{TTmap}
\hat\CC_{0(0,0)}\longrightarrow \hat\CC_{0(0,0)}^{\rm Free}~,
\end{equation}
where the multiplet on the righthand side is the stress tensor multiplet of free super-Maxwell theory \footnote{As in the footnote below \eqref{EDmap}, we should subtract lower-dimensional multiplets to make the map precise (in particular, a real linear combination of $\bar{\CE}_{6/5}$ and $\CE_{-6/5}$). This shift will not affect the Schur operator (discussed below) in the IR since free vector multiplets lack $R=1$ states.}.

Both multiplets appearing on the righthand side of \eqref{EDmap} and \eqref{TTmap} are \lq\lq Schur" multiplets \cite{Gadde:2011uv}. The corresponding highest-$SU(2)_R$ weight states (with highest Lorentz weight) are \lq\lq Schur" operators. These operators, along with $\partial_{+}:=\partial_{+\dot+}$, form an interesting ring of operators in 4d we will refer to as the \lq\lq Schur" ring and constitute the quarter-BPS observables we mentioned above. In the case of \eqref{TTmap}, the Schur operator map is
\begin{equation}\label{TTmapS}
\hat\CC_{0(0,0)}\ni J:=J^{11}_{+\dot+}\longrightarrow \lambda^1_+\bar\lambda^1_{\dot+}\in \hat\CC_{0(0,0)}^{\rm Free}~,
\end{equation}
where $J^{11}_{+\dot+}$ is the highest-weight UV $SU(2)_R$ current, and $\lambda^1_+\in\bar\CD^{\rm Free}_{0(0,0)}$, $\bar\lambda^1_{\dot+}\in\CD^{\rm Free}_{0(0,0)}$ are IR gauginos.

The MAD Schur ring only has $\hat\CC_{R(j,j)}$ multiplets \cite{Buican:2021elx}. Moreover, it has an \lq\lq extremal" subsector. These are Schur operators and multiplets that, for a given $SU(2)_R$ weight, $R$, have lowest spin, $j$. The stress tensor multiplet is the case $R=j=0$. More generally, extremal Schur operators, $\CO_{R, {\rm Ext}} \in\hat\CC_{R(\frac12 R(R+2) ,\frac12R(R+2) )} $, map as follows \cite{Buican:2021elx}
\begin{eqnarray}\label{Ext}
\CO_{R,{\rm Ext}}      &\longrightarrow& 
 \Big(\lambda^1_+\partial_+\lambda^1_+\cdots\partial^R_+\lambda^1_+\Big) \Big (\bar\lambda^1_{\dot+}\partial_+\bar\lambda^1_{\dot+}\cdots\partial^R_+\bar\lambda^1_{\dot+} \Big)
 \cr&&
  \sim \lambda^1_+\bar\lambda^1_+\partial^2\Big(\lambda^1_+\bar\lambda^1_{\dot+}\Big)\cdots\partial^{2R}\Big(\lambda^1_+\bar\lambda^1_{\dot+}\Big)~, \ \ \ \
\end{eqnarray}
where we have used Fermi statistics to rearrange the gauginos in a fashion of use below.

Given this discussion, it is natural to expect a general relation between the UV and IR Schur rings. However, there are potential obstacles: {\bf(a)} All IR Schur operators need not come from UV Schur operators. {\bf(b)} In general SCFTs, UV Schur operators can decouple along flows to the Coulomb branch.

Regarding {\bf(a)}, \eqref{EDmap} implies the UV origin of the gauginos is in the MAD chiral sector, not the Schur sector \footnote{$\bar{\CE}_{6/5}$ is not Schur.}. Moreover, because the IR is free, it has higher spin symmetries which are absent in the UV \cite{Maldacena:2011jn,Alba:2015upa}. The breaking of these symmetries in the flow back to the UV is encoded as follows \cite{Bhargava:2022yik}
\begin{equation}\label{ComplexHS}
\bar\CC_{0,7/5(k,k-1)}\longrightarrow\hat\CC_{0(k,k-1)}^{\rm Free}~,\ \ \ k=1,2,\cdots~.
\end{equation}
On the righthand side, we have emergent complex higher spin current multiplets, while, on the lefthand side, we have \lq\lq longer" protected multiplets that include non-vanishing divergences of would-be MAD higher-spin currents. For real higher-spin currents \cite{Bhargava:2023}
\begin{equation}\label{RealHS}
\CA^{\Delta}_{0,r(k,k)}\longrightarrow\hat\CC_{0(k,k)}^{\rm Free}~,\ \ \ k=1/2,1,\cdots~.
\end{equation}
On the lefthand side, we have certain UV long multiplets. Therefore, a main task is to carve out the subsector of IR Schur operators corresponding to UV Schur operators. This discussion is summarized in Fig. \ref{Flowdiag}.

Regarding {\bf(b)}, note it is common for Schur operators to decouple in Coulomb branch flows. For example, on a genuine Coulomb branch consisting of free vectors at generic points, flavor symmetries decouple. Since flavor symmetry Noether currents lie in Schur multiplets, Schur operators can decouple. More generally, decoupling is unrelated to flavor.

Given the \lq\lq closeness" of the MAD SCFT to the Coulomb branch, it is reasonable to expect both obstacles are irrelevant. We will soon see this is the case.

A useful feature of the UV Schur ring is its simplicity. Indeed, as explained in the Supplemental Material, it is generated by the $\partial_+^iJ$ subject to
\begin{equation}\label{JJnull}
\hat\CC_{1(1/2,1/2)}\ni J^2:= \ :JJ:\ =0~,
\end{equation}
where \lq\lq:$\cdots$:" denotes the normal-ordered product \footnote{Since there are no corresponding OPE singularities, normal ordering does not involve subtractions.}.

Consistency with \eqref{TTmap} suggests looking for an IR null state related to \eqref{JJnull}. Indeed, using \eqref{TTmap}, the non-trivial UV dynamics leading to \eqref{JJnull} maps to an IR constraint enforced by Fermi statistics \footnote{We therefore have a physical interpretation of the \lq\lq fermionic" constraints discussed in \cite{Foda:2019guo}.}
\begin{equation}\label{FermiConstraint}
\hat\CC_{1(1/2,1/2)}\ni J^2\longrightarrow (\lambda^1_+\bar\lambda^1_{\dot+})^2=0\in\hat\CC_{1(1/2,1/2)}^{\rm Free}~.
\end{equation}

Given this discussion, we propose the following map:

\noindent
{\bf Main statement:} An arbitrary monomial in the MAD Schur ring is mapped as follows to the IR
\begin{eqnarray}\label{proposal}
\CS_{\rm MAD} \ni \partial^{i_1}_+J\cdots\partial_+^{i_n}J
&\longleftrightarrow&  \partial^{i_1}_+(\lambda^1_+\bar\lambda^1_{\dot+})\cdots\partial^{i_n}_+(\lambda^1_+\bar\lambda^1_{\dot+})
\cr& &\in
\tilde\CS_{\rm Free\ Vector}\subset\CS_{\rm Free\ Vector}~.\nonumber \\
\end{eqnarray}
\noindent
Here $\CS_{\rm Free\ Vector}$ is the set of all IR Schur operators \footnote{These operators correspond to arbitrary combinations of gauginos and derivatives.}. 
An important feature of \eqref{proposal} is that individual gauginos and higher-spin currents are not in the map's target (i.e., \eqref{proposal} is consistent with \eqref{EDmap}, \eqref{ComplexHS}, and \eqref{RealHS}).

On the other hand, Fermi statistics naively looks more constraining than \eqref{FermiConstraint}. Indeed, $(\lambda^1_+)^2=(\bar\lambda^1_{\dot+})^2=0$ implies \eqref{FermiConstraint}, not vice versa. Therefore, we should make sure there are as many null states on one side of \eqref{proposal} as on the other.

Using results on \lq\lq leading ideals," \cite{greuel2008singular,bruschek2011arc} we will show that, for operators in \eqref{proposal}, Fermi statistics is equivalent to \eqref{FermiConstraint}. Combined with the fact that the $\partial^i_+J$ subject to \eqref{JJnull} generate the UV Schur ring, we establish \eqref{proposal}. As a byproduct, we show that the Macdonald index, an observable that counts Schur operators, is exactly computable in the IR.

We have avoided discussing the relation of 4d Schur rings to 2d vertex operator algebras (VOAs) \cite{Beem:2013sza}. The main reason is our discussion is inherently 4d, and the twisting in \cite{Beem:2013sza} somewhat obscures this (we will return to the 2d free field construction of \cite{Beem:2019tfp,Bonetti:2018fqz} in section \ref{disc}). However, as we discuss, the 4d/2d map is useful in deriving \eqref{JJnull}. Moreover, results on arc spaces \cite{arakawa2019singular} imply the UV Schur ring is characterized as claimed around \eqref{JJnull} \cite{bai2020quadratic}.

The plan of the paper is: in section  \ref{SchurMAD} we briefly review the MAD theory and its Schur sector. In section \ref{ideal}, we show Fermi statistics does not lead to additional constraints spoiling \eqref{proposal}. We conclude with a general discussion in section \ref{disc}. 

\section{The MAD theory's Schur sector}\label{SchurMAD}
We briefly review the construction of the Schur ring, describe its counting by the Macdonald index, and discuss the example of the MAD theory. Finally, we explain how the map in \cite{Beem:2013sza} can be used to derive \eqref{JJnull} and explain how the UV Schur ring is generated (details appear in the Supplemental Material).

A Schur operator, $\CO$, satisfies
\begin{equation}\label{constraints}
\left\{\CQ^1_-,\CO\right]=\left\{\tilde\CQ_{2\dot-},\CO\right]=\left\{\CS_1^-,\CO\right]=\left\{\tilde S^{2\dot-},\CO\right]=0~.
\end{equation}
Numerical indices are $SU(2)_R$ weights, and signs indicate spin weights. These relations imply
\begin{equation}
E_{\CO}=2R_{\CO}+j_{\CO}+\bar j_{\CO}~,
\end{equation}
where the lefthand side is the scaling dimension, $R$ is the $SU(2)_R$ weight, $r$ is the $U(1)_r$ charge, and $j$, $\bar j$ denote spin weights. Operators carrying these quantum numbers are counted by the Macdonald index
\begin{equation}
\CI_M(q,T):={\rm Tr}(-1)^Fq^{E-R}T^{R+r}~,
\end{equation}
where the trace is over the space of Schur operators, $q$ and $T$ are fugacities, and $(-1)^F$ is fermion number.

The MAD Macdonald index was computed via TQFT in \cite{Song:2015wta}, but the elegant expression in \cite{Foda:2019guo} is particularly useful
\begin{equation}\label{MADM}
\CI^{\rm MAD}_M(q,T)=\sum_{n=0}^{\infty}{q^{n^2+n}\over(q)_n}T^n~, \ \ \ (q)_n:=\prod_{i=1}^n(1-q^i)~.
\end{equation}
Here, $\partial_+^iJ$ contributes $q^{i+2}T$, and products of operators give products of contributions.

To interpret the physical states contributing to \eqref{MADM}, we briefly recall the Schur ring to VOA map \cite{Beem:2013sza} (see \cite{Beem:2013sza} for further details). The idea is to perform an $SU(2)_R$ twist of right-moving $sl(2,\mathbb{R})$ transformations on a plane inside $\mathbb{R}^4$. Then, the algebraic constraints in \eqref{constraints} imply that Schur operators are non-trivial cohomology elements
\begin{eqnarray}\label{Schurcond2}
\left\{\mathds{Q}_i,\mathcal{O}(0)\right]&=&0~, \ \ \ \mathcal{O}(0)\ne\left\{\mathds{Q}_i,\mathcal{O}'(0)\right]~,\cr \mathds{Q}_1&:=&\CQ^1_-+\tilde S^{2\dot-}~, \ \ \ \mathds{Q}_2:=S_1^{-}-\tilde Q_{2\dot-}~.
\end{eqnarray}
Moreover, twisting guarantees that planar translations by $\partial_{-\dot+}$ are cohomologically trivial while those generated by $\partial_+$ are not. As a result, we can map twisted-translated $\mathds{Q}_i$ cohomology classes in \eqref{Schurcond2} to operators that only depend on a holomorphic planar coordinate, $z$. These latter operators are members of a VOA. Particularly relevant for us are the maps
\begin{eqnarray}\label{4d2d}
\chi([J]_{\mathds{Q}})&=&T_{2d}~,\ \ \ \chi(\partial_{+})=\partial_z:=\partial~,\ \ \ c_{2d}=-12c_{4d}~, \cr h&=&E-R~,
\end{eqnarray}
where $[J]_{\mathds{Q}}$ is the cohomology class of the $SU(2)_R$ current \footnote{We take $\mathds{Q}_i\to\mathds{Q}$, since the cohomology is independent of $i$.}, $T_{2d}$ is the VOA stress tensor, $c_{2d}$ is the corresponding central charge (twisting leads to 2d non-unitarity), $h$ is the holomorphic scaling dimension, and $\chi$ is the 4d/2d map.

Using this construction (specifically the $T\to1$ limit of \eqref{MADM}, which becomes the VOA vacuum character) and some orthogonal arguments we will return to in the discussion section, the authors of \cite{Cordova:2015nma} argued that the VOA corresponding to the MAD theory is the Lee-Yang vacuum module \footnote{Note that the central charge of the MAD theory gives the correct $c_{2d}=-22/5$ via \eqref{4d2d}.}
\begin{equation}\label{MADchir}
\chi({\rm MAD})={\rm Vir}_{c_{2d}=-22/5}~.
\end{equation}
This VOA is built from normal-ordered products of $\partial^iT_{2d}$ for arbitrary $i$.

Famously, Lee-Yang has an $h=4$ null state
\begin{equation}\label{LYnull}
T_{2d}^2-{3\over10}\partial^2T_{2d}=0~,\ \ \ T^2_{2d}:=\ :T^2_{2d}:~.
\end{equation}
This null relation is the 2d incarnation of \eqref{JJnull} (e.g., see \cite{Liendo:2015ofa}). Indeed, from the general construction in \cite{Beem:2013sza}, we can work out the terms that do not vanish in the $z\to0$ limit of the $T_{2d}(z)T_{2d}(0)$ OPE by considering all $SU(2)_R$ components of the 4d $J_{+\dot+}^{i_1i_2}(z)J_{+\dot+}^{j_1j_2}(0)$ OPE (recall $J:=J^{11}_{+\dot+}$) and looking for Schur operators with $h\le4$.

In particular, the null state in \eqref{LYnull} corresponds to a 4d null state with $h=4$, and multiplet selection rules imply this operator has $R=2$ \footnote{These selection rules imply that it is the Schur operator in a multiplet of type $\hat{\CC}_{1(1/2,1/2)}$ (e.g., see \cite{Liendo:2015ofa,Kiyoshige:2018wol})}. It therefore corresponds to the vanishing normal-ordered product
\begin{equation}\label{JJ4d}
J(z)J(0)\supset J^2(0)=0~.
\end{equation}
This equation is a non-trivial UV dynamical constraint.

Given that the VOA in \eqref{MADchir} is strongly generated by $T_{2d}$, it is natural to conjecture that the 4d Schur ring is generated by normal-ordered products of $\partial^iJ$ subject to \eqref{JJ4d} \footnote{This statement is nontrivial. Indeed, the number of generators in 4d and 2d is generally different. For example, when the 2d theory has a Sugawara stress tensor, it has one fewer generators relative to 4d \cite{Beem:2013sza} (see \cite{Buican:2015ina} for closely related theories).}. Let us call this ring $R^{\rm MAD}_{\infty}$ and define
\begin{equation}\label{Rinf}
R^{\rm MAD}_{\infty}:=\mathbb{C}[J,\partial_+ J, \partial_+^2J,\cdots]/\langle J^2 ,\partial_+(J^2 )~,\cdots\rangle~.
\end{equation}

Indeed, as explained in the Supplemental Material, recent results on arc spaces imply the counting of operators in $R^{\rm MAD}_{\infty}$ matches \eqref{MADM}. More precisely,
\begin{equation}\label{HSM}
{\rm HS}_{R^{\rm MAD}_{\infty}}(q,T) =\CI_M^{\rm MAD}(q,T)~,
\end{equation}
where the lefthand side is the Hilbert series of $R^{\rm MAD}_{\infty}$. Since all operators involved are bosonic, this result is a highly non-trivial check of the claim that the 4d Schur ring is generated by products of $\partial^iJ$ subject to \eqref{JJ4d}.

Next we apply the RG map \eqref{TTmapS} and reproduce the Macdonald index in terms of the IR degrees of freedom and Fermi statistics.

\section{IR Fermi statistics}\label{ideal}
When flowing to the IR, $J\to\lambda^1_+\bar\lambda^1_{\dot+}$, and, as explained around \eqref{FermiConstraint}, the UV dynamics that lead to \eqref{JJ4d} manifest as IR Fermi statistics. Therefore, our goal is to apply the map in \eqref{proposal} and reproduce \eqref{HSM} in the IR.

Therefore, we must show Fermi statistics doesn't imply additional constraints. Intuitively, we expect this not to be an issue since the IR operators we consider do not probe the full emergent Schur ring. For example, they are blind to accidental higher-spin symmetries.

To make our discussion precise, we first write a UV basis of operators and make contact with the extremal Schur operators \eqref{Ext}. As explained in the Supplemental Material, we can use results in algebraic geometry to show that a suitable basis consists of
\begin{eqnarray}\label{Jbases}
&&\p_+^{n_1} J \; \p_+^{n_2 } J\cdots \p_+^{n_k} J~, \ \ \  
0\le n_1<n_2<\cdots <n_k~, \cr&& n_{i+1} -n_i \ge 2~, \ \sum_{i=1}^k n_i=n~, \ \ \ n\in\mathbb{Z}_{\ge0}~.
\end{eqnarray}
The extremal case  \eqref{Ext} has $n_{i+1}-n_i=2$ and $n_1=0$.

Applying the RG map \eqref{TTmapS} to \eqref{Jbases}, we get a composite operator made of fermions. Due to Fermi statistics, which is generally stronger than  \eqref{JJ4d}, one may worry the operator vanishes.  To show it does not, we pick a representative non-vanishing term. In general, there are multiple non-vanishing terms after distributing derivatives. We simply should make a consistent choice. To that end, we choose
\begin{eqnarray}\label{lambdaexpr}
&&\p_+^{m_1}\lambda^1_+\p_+^{m_1'}\bar\lambda^1_{\dot+} \; \;
\p_+^{m_2}\lambda^1_+\p_+^{m_2'}\bar\lambda^1_{\dot+} \; \;
 \cdots\;
\p_+^{m_k}\lambda^1_+\p_+^{m_k'}\bar\lambda^1_{\dot+} ~,\cr
&& m_i'-m_i=0~,\ 1~, \ \ \  0\le m_1<m_2 <\cdots < m_k~,\cr&&0\le m_1'< \cdots < m_k'~.
\end{eqnarray}
Clearly there is a one-to-one correspondence ({\it not} equality) between \eqref{lambdaexpr} and \eqref{Jbases} after setting $m_i+m_i'=n_i, m_i=\lfloor{\frac{n_i}{2}\rfloor}$, and $m_i'=\lfloor{\frac{n_i+1}{2}\rfloor}$. At the level of operators,
 \begin{eqnarray}
&&\p_+^{n_1} J \;  \cdots \p_+^{n_k} J 
\\&\longleftrightarrow &
 \Big(\p^{\lfloor{\frac{n_1}{2}\rfloor}}_+\lambda^1_+\p^{\lfloor{\frac{n_1+1}{2}\rfloor}}_+\bar\lambda^1_{\dot+}\Big) \; \;
 \Big(\p_+^{\lfloor{\frac{n_2}{2}\rfloor}}\lambda^1_+\p_+^{\lfloor{\frac{n_2+1}{2}\rfloor}}\bar\lambda^1_{\dot+}\Big) \;\cr&&
 \cdots\; \Big(\p_+^{\lfloor{\frac{n_k}{2}\rfloor}}\lambda^1_+\p_+^{\lfloor{\frac{n_k+1}{2}\rfloor}}\bar\lambda^1_{\dot+}\Big)~.
\end{eqnarray}
In particular, since $n_{i+1} -n_i \ge 2$, we see $ {\lfloor{\frac{n_i}{2}\rfloor}} \neq   {\lfloor{\frac{n_j}{2}\rfloor}}$ as long as $i\neq j$. Therefore, the fermions do not annihilate. It is also obvious that operators in \eqref{lambdaexpr} are linearly independent for different sets of $m_i,m_i'$.

As a result, \eqref{Jbases} gives an IR basis. We see that Fermi statistics does not over-constrain our subring of observables, and we reproduce \eqref{HSM} in the IR.

\section{Discussion}\label{disc}
\begin{figure}[h!]
\centering
\includegraphics[width=3.5in]{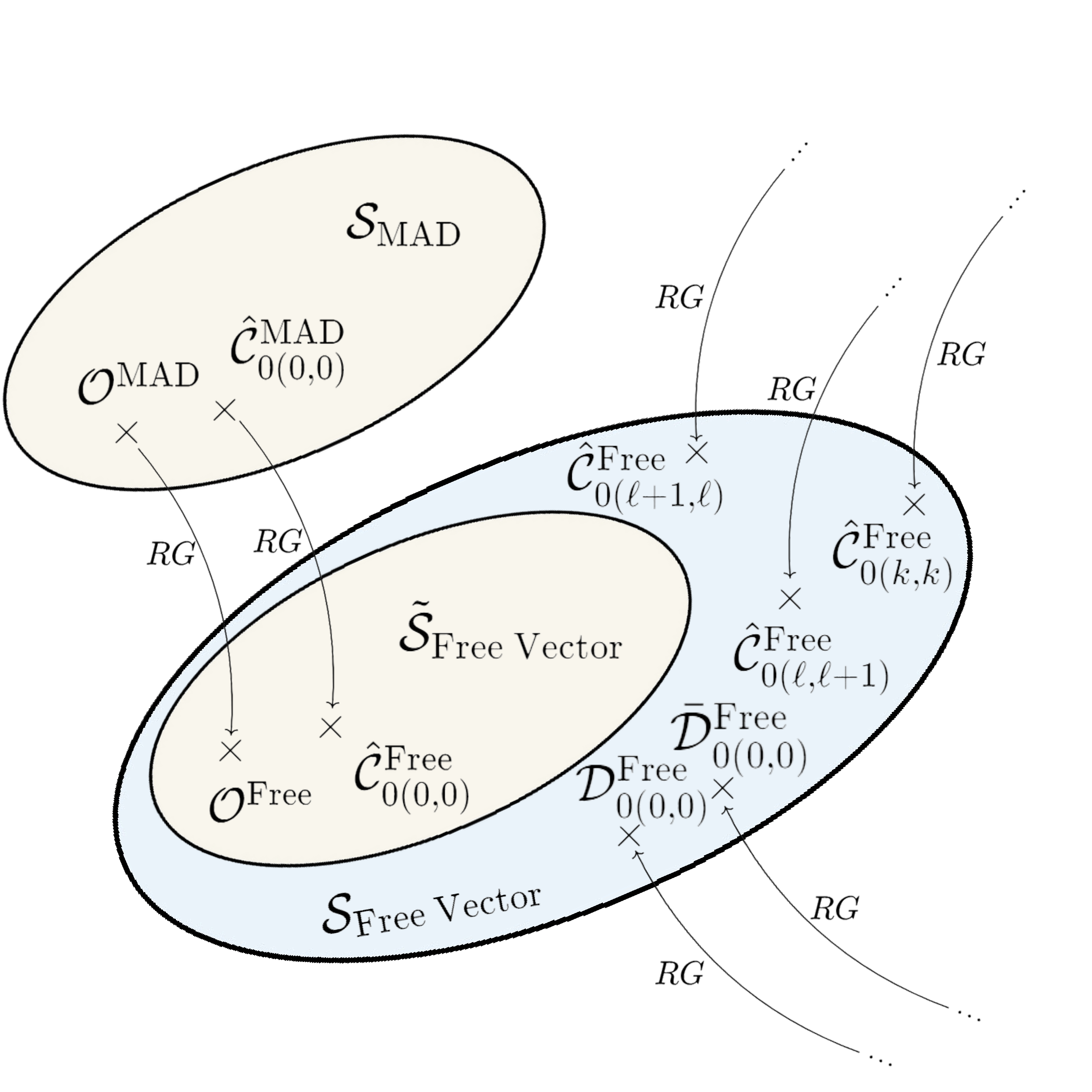}
\caption{RG maps to the IR Schur sector, $\CS_{\rm Free\ Vector}$. We describe the flow between the UV MAD Schur sector, $\CS_{\rm MAD}$, and a closed subsector of the IR Schur operators, $\tilde\CS_{\rm Free\ Vector}$ (yellow shading). IR Schur operators in the complement of $\tilde\CS_{\rm Free\ Vector}$ (blue shading) come from non-Schur UV operators.}\label{Flowdiag}
\end{figure}

As far as we are aware, \eqref{proposal} is the first exact map of non-chiral quarter-BPS observables along a strongly coupled RG flow. UV dynamics giving rise to null relations are reduced to IR Fermi statistics (it would be interesting to derive these relations via UV defect endpoint operators). Noting that the IR gauginos are related by supersymmetry to the coordinates on the Coulomb branch, and thinking of an arc space as an infinite-dimensional generalization of a tangent space, we see that our results constitute a certain geometrical completion of Seiberg-Witten theory for the MAD SCFT.

It is surprising that a Coulomb branch flow knows so much about the Schur sector (this sector is typically associated with the Higgs branch). At the same time, this fact strengthens our conjecture \eqref{CConj} and shows that Coulomb branch and Schur sector physics unify into a deeper structure (see also \cite{Buican:2015hsa,Fredrickson:2017yka}).

The above phenomena are indirectly related to those in \cite{Cordova:2015nma}. There the authors computed a less refined limit of the superconformal index by summing over massless and massive Coulomb branch BPS states. We instead keep track of the Schur operators along the RG flow. In so doing, we recover additional 4d quantum numbers ($SU(2)_R$ charges).

When does the above construction generalize to other Coulomb branch flows? A reasonable conjecture is that it generalizes whenever the UV \lq\lq hidden" symmetries of the Schur ring (Virasoro here) are all related to symmetries of the full 4d theory that are not explicitly broken along the RG flow and do not decouple ($SU(2)_R$ in the present case). Indeed, as we show in the Supplemental Material, $(A_1,A_{2r})$ SCFTs have similar IR embeddings of their Schur sectors. These theories have purely Virasoro hidden symmetry related to unbroken $SU(2)_R$.

On the other hand,  consider Coulomb branch flows for theories with $W_{N>2}$ symmetry. For example, the $(A_2, A_3)$ SCFT has (hidden) $W_3$ symmetry \cite{Cordova:2015nma}. Using the Macdonald index \cite{Song:2015wta,Foda:2019guo}, it is easy to argue that the $W_3$ current sits in a $\hat\CC_{1(0,0)}$ multiplet. It is simple to check that the corresponding Schur operator cannot be built from gauginos and derivatives. In this case, we expect the $W_3$ symmetry to decouple along flows to generic points on the Coulomb branch \footnote{Unlike Virasoro, it is unclear whether this $W_3$ is related to a 4d symmetry.}

Let us also discuss how our work is related to known free field constructions \cite{Beem:2019tfp,Bonetti:2018fqz}. There the authors studied Higgs branch RG flows and focused on massless degrees of freedom (in $\CN>2$ SUSY, such moduli spaces embed in larger structures that include Coulomb branches). In these cases, some of the symmetries are spontaneously broken, but one can construct UV 2d VOA operators in terms of IR 2d VOA degrees of freedom (see also related work in \cite{Adamovic:2019tcl}) \footnote{Presumably it is important that the breaking is spontaneous and not explicit.}.

We have instead followed 4d operators along Coulomb branch RG flows. Understanding such flows from the Schur sector perspective is crucial, since the Coulomb branch is the most universal moduli space of an interacting 4d $\CN=2$ SCFT \footnote{Indeed, all our examples have trivial Higgs branches.}. A more closely related 2d version of our discussion in the spirit of \cite{Beem:2019tfp,Bonetti:2018fqz} is to fermionize the Coulomb gas construction of the Lee-Yang theory (along the lines of \cite{Bonora:1989wk,Feigin:1981st}). However, this would require us to express the IR version of the UV stress tensor as a composite not built purely out of 2d avatars of IR gauginos (see (2.1) of \cite{Bonora:1989wk}) \footnote{Perhaps such a construction allows one to reconstruct $W_{N>2}$ minimal model VOAs from the Coulomb branch.}.

As emphasized in \eqref{EDmap}, \eqref{ComplexHS}, \eqref{RealHS}, and Fig. \ref{Flowdiag}, the full IR Schur sector is connected via RG flow to various UV sectors. It will be interesting to use these maps to further constrain the UV (from our Coulombic generation conjecture, we expect the corresponding UV operators generate the MAD theory). For example, we can consider products of operators in \eqref{proposal} with other operators and infer aspects of the $\bar\CC$ spectrum \cite{Bhargava:2023}.

Finally, it is tempting to take our results and search for a geometrical completion of Seiberg-Witten theory in more general 4d $\CN=2$ QFTs.

\acknowledgements{We thank A.~Banerjee, S.~Gukov, A.~Hanany, T.~Nishinaka, and S.~Wood for enlightening discussions. We also thank S.~Wood for initial collaboration on this work and T.~Nishinaka for collaboration on closely related work \cite{Buican:2021elx}. M. B. and H. J. were partially supported by the grant “Relations, Transformations, and Emergence in Quantum Field Theory” from the Royal Society and the grant “Amplitudes, Strings and Duality” from STFC. C. B. was partially supported by funds from QMUL.}

\bigskip\bigskip\bigskip\bigskip

\begin{appendix} 

\appendix*{\bf Supplemental Material I: Macdonald index, arc space and the 4D theory}\label{arc}

We begin by briefly reviewing the Macdonald index. This observable counts operators that obey the relations in (11) and (12) of the main text. These operators sit in $\hat{\mathcal{B}}_R$, $\mathcal{D}_{R,(0,\bar{j})}$, $\bar{\mathcal{D}}_{R,(j,0)}$, and $\hat{\mathcal{C}}_{R(j,\bar{j})}$ multiplets in the nomenclature of \cite{Dolan:2002zh}. One can often find a closed-form expression for the Macdonald index. For example, in the $\mathcal{N} = 2$ free massless hypermultiplet, one finds the following freely generated answer
\begin{equation}
    \CI_M^{\rm Free\ hyper} = {1\over(z \sqrt{t},q)_{\infty} (\frac{\sqrt{t}}{z},q)_{\infty}}~,
\end{equation}
where $(a,q)_n := \prod_{k=0}^{n-1}(1 - a q^k)$, and $z$ is a flavour fugacity.

Next, our goal is to explain the equality in (21) of the main text. For ease of reference, we reproduce it here
\begin{equation}\label{HSMapp}
{\rm HS}_{R^{\rm MAD}_{\infty}}(q,T) =\CI_M^{\rm MAD}(q,T)~.
\end{equation}
Recall from the discussion below (14) of the main text that $\partial_+$ contributes $q$, and $J$ contributes $q^2T$ to the righthand side of this equation. The lefthand side is the Hilbert series that counts operators in $R_{\infty}^{\rm MAD}$. We will give a precise definition of this quantity below.

If \eqref{HSMapp} holds, then $\CI^{\rm MAD}_M$ can be obtained from words built out of $J$ and the derivative $\p_+$, subject to the condition $J^2=0$. This is because $R^{\rm MAD}_{\infty}$ defined in (20) of the main text contains all such operators. Therefore, we would like to check whether
 \begin{eqnarray}\label{check}
\CI^\text{MAD}_M(q,T)&=&\sum_{k=0}^\infty \frac{q^{k^2+k}}{(q)_k} T^k
\nonumber \\ &\stackrel{?}{=}&
\sum_{n,k=0}^\infty  \dim V_{n,k} \; (q^2 T)^k q^n
\cr&=&
\sum_{n,k=0}^\infty  \dim V_{n,k} \;   q^{2k+n} T^k~.
 \end{eqnarray}
Here $V_{n,k}$ is the set of operators built from $k$ $J $'s and $n$ derivatives $\p_+$ (subject to $J^2=0$), which, in general, take the form (we have suppressed complex coefficients in front of each term in the sum for simplicity)
 \be\label{Jexpr}
 \sum_{n_1, \cdots, n_k=0\atop \sum n_i =n}^\infty  \p^{n_1} J \; \p^{n_2 } J\;  \cdots \p^{n_k} J~,
 \ee
and  $\dim V_{n,k}$ is the dimension of each such linearly independent subspace.
 
 As a result, to prove \eqref{HSMapp}, we need to show that 
   \begin{eqnarray} \label{J2Mac}
 \sum_{n,k=0}^\infty\!\!\!  \dim V_{n,k} \;   q^{ n}  p^k   \!\! 
 &  \stackrel{?}{= }&
\sum_{k=0}^\infty \frac{q^{k^2-k}}{(q)_k} p^k~,   \ \  (p=q^2 T)~.\qquad
 \end{eqnarray}
 Interestingly, the RHS can be identified with a $q$-hypergeometric series \footnote{For more details, see \url{https://en.wikipedia.org/wiki/Basic_hypergeometric_series}.}
 \begin{eqnarray} \label{qhyper}
\sum_{k=0}^\infty \frac{q^{k^2-k}}{(q)_k} p^k 
&=& 
 {}_r\phi_{r+1}\Big[{   {a_1, \cdots , a_r} \atop{ {a_1, \cdots , a_r, 0}} }; q,p \Big]  
\cr &=&
1+p+p q+\left(p+p^2\right) q^2
\cr & &
+\left(p+p^2\right) q^3
+\left(p+2 p^2\right) q^4+\cdots ~. \nonumber\\
 \end{eqnarray}
 for arbitrary $r$ and $a_i$, where we have also explicitly written the perturbative expansion for the first few orders.  
 
It is easy to compute $ \dim V_{n,k}$ numerically to high order and verify this statement. Below, we will also prove it analytically.

To do so, it is useful to introduce the concept of an arc space. An arc space is a special kind of topological space that is intimately connected with the singularities of algebraic varieties. In the context of QFT, such spaces have appeared in various places (e.g., see \cite{Beem:2017ooy,Beem:2019tfp,arakawa2019singular}). In our case, the arc space encodes the operators in the Schur sector, and one can characterize the Schur spectrum from the associated arc space Hilbert series. 

Here we follow \cite{arakawa2019singular} and start with the affine scheme 
 \be
 X=\text{Spec } R~, \ \ \ R=\mathbb C[x_1, \cdots, x_N]/ \langle{f_1, \cdots f_l\rangle}~,
 \ee
 where $f_i \in \mathbb C[x_1, \cdots, x_N]$ are polynomial relations.
 
 From this structure, we have the jet scheme, $X_m$, which can be thought of as a generalization of the notion of a tangent space. It is given by 
\begin{eqnarray}
 X_m&=&\text{Spec } R_m~, \ \ \cr  R_m&=& \mathbb C[x^{(i)}_1, \cdots, x^{(i)}_N]/ \langle{f^{(i)}_1, \cdots f^{(i)}_l\rangle}~,  0\le i \le m~.\qquad\quad
\end{eqnarray}
In writing the above ideals, we introduced a derivation, $D$, such that $D(x^{(i)}_j) = x^{(i+1)}_j $ if $0 \le i <m$ and $D(x^{(i)}_j) = 0$ if $i=m$. This definition then specifies the action of $D$ on all $\mathbb C[x^{(i)}_1, \cdots, x^{(i)}_N]$. In particular, $  f^{(i )}_j:=D^i(f_j) $ is also a polynomial.

Given this discussion, we can consider the inverse limit and obtain the arc space
\begin{eqnarray}\label{asdef}
X_\infty &=&\lim\limits_{  \leftarrow} X_m \simeq\text{Spec } R_\infty~, \cr 
R_\infty &=&\mathbb C[x^{(i)}_1, \cdots, x^{(i)}_N]/ \langle{f^{(i)}_1, \cdots f^{(i)}_l\rangle}~,\ \ \  i\ge 0~.\ \ \ \ \ \ \ 
\end{eqnarray}

In \cite{Beem:2017ooy,Beem:2019tfp,arakawa2019singular}, the above construction arises in the context of 2d VOAs, and $R=\mathcal{R}_{\mathcal{V}}$ is the associated Zhu's $C_2$ algebra. Roughly speaking, this is a commutative 2d algebra obtained by getting rid of all operators containing derivatives in the VOA $\mathcal{V}$
\begin{equation}
\mathcal{R}_{\mathcal{V}}=\mathcal{V}/\mathcal{C}_2(\mathcal{V})~,\ \ \ \mathcal{C}_2:={\rm Span}\left\{a_{-h_a-1}b|a,b\in\mathcal{V}\right\}~.
\end{equation}
When the 4d SCFT has a Higgs branch, Zhu’s  $C_2$ algebra enables one to reconstruct this moduli space \cite{Beem:2017ooy}. In the case of the MAD theory, there is no Higgs branch. However, Zhu's $C_2$ algebra still contains important information about this theory. Indeed, from (18) of the main text, it is easy to see that
\begin{equation}\label{MADRV}
\mathcal{R}_{{\rm Vir}_{c=-22/5}}=\mathbb{C}[x]/\langle x^2\rangle~.
\end{equation}

Constructing the arc space associated with \eqref{MADRV} and showing that its operators are counted as in \eqref{check} is strong evidence for the fact that the arc space undoes the twisting of the MAD theory that led to the Lee-Yang VOA. At a physical level, the arc space therefore provides an inverse map from 2d to 4d for the case at hand (and also for the generalizations we discuss in section III of the Supplemental Material).

To prove \eqref{check}, we consider $N=1$ and $x_1=J$ in \eqref{asdef}. The derivation, $D$, can be regarded as the derivative acting on local operators. Then we can identify $x_1^{(i)} =\p_+^i J$. The ideal is generated by $f_1^{(0)}=J^2$ and, more generally, all $f_1^{(i)}=\p_+^i (J^2)$.  As a result, the above arc space is exactly the space describing operators made out of $J$ and derivatives   in \eqref{Jexpr} subject to the constraint $J^2=0$. In other words, $R_{\infty}\to R_{\infty}^{\rm MAD}$, and we need to consider 
\be\label{arcspace}
X_\infty^{\rm MAD}  =\text{Spec } R^{\rm MAD}_\infty, \ \ \ 
R^{\rm MAD}_\infty =\mathbb C[x^{(i)}  ]/ \langle{(x^2)^{(i)} \rangle}, \ \ \  i\ge 0~,
\ee
or, as described around (20) of the main text using physical operators
\be\label{Rinfapp}
R_\infty^{\rm MAD} =\mathbb C[J, \p_+ J, \p_+^2 J, \cdots ] /\langle{J^2 , \p_+ (J^2) , \cdots\rangle}~.
\ee
This ring is bi-graded, and we can assign weights $(i,1)$ to $x^{(i)} =\p^i_+ J$. These weights correspond to the $(E-3R,R)$ quantum numbers of operators in 4d.

To complete the proof, we need to first define the associated Hilbert series
\be\label{HSdef}
\text{HS}_{R_{\infty}^{\rm MAD}} (q,p):=\sum_{n,k=0}^\infty\dim (R^{\rm MAD}_\infty)_{n,k} \; q^n p^k~,
\ee
where $\dim (R^{\rm MAD}_\infty)_{n,k}$ is the dimension of the subring with weight $(n,k)$. As a result, we have $\dim (R^{\rm MAD}_\infty)_{n,k}=\dim V_{n,k}$ for the number of such linearly independent operators.

The remaining goal is to compute the Hilbert series in \eqref{HSdef}. Fortunately, this has been done in \cite{bai2020quadratic}. Indeed, from (7.1) in that reference, we learn that 
\begin{eqnarray}\label{HSinf}
\text{HS}_{R^{\rm MAD}_\infty} (q,p)&=&\sum_{n,k=0}^\infty\dim (R_\infty)_{n,k} \; q^n p^k
\cr&=&\sum_{n,k=0}^\infty \dim V_{n,k} \; q^n p^k
\cr&=&\sum_{k=0}^\infty \frac{q^{k^2-k} }{(q)_k} p^k~.
\end{eqnarray}
This discussion thus proves the identity \eqref{J2Mac}. Therefore, operators made out of $J$ and derivatives in \eqref{Jexpr} subject to the constraint $J^2=0$ reproduce all the Schur operators and (upon the substitution $p\to q^2T$)  the associated Macdonald index as in \eqref{HSMapp}.

\appendix*{\bf Supplemental Material II: Leading ideals and a basis for the arc space}

In this section, our goal is to show that the states in (22) of the main text form a basis for $R_{\infty}^{\rm MAD}$. For ease of reference, we reproduce these states here
\begin{eqnarray}\label{Jbasesapp}
&&\p_+^{n_1} J \; \p_+^{n_2 } J\cdots \p_+^{n_k} J~, \ \ \  
0\le n_1<n_2<\cdots <n_k~, \cr&& n_{i+1} -n_i \ge 2~, \ \sum_{i=1}^k n_i=n~, \ \ \ n\in\mathbb{Z}_{\ge0}~.
\end{eqnarray}
For a given $k$ and $n$, we wish to show that the above states form a basis for the space $V_{n,k}$.

To prove this statement, we first note that, following theorem 6.3 of \cite{bruschek2011arc} (see also proposition (5.2) of that reference), we have
\be\label{ILI}
\text{HS}_{\mathbb C[x,\cdots]/I} =\text{HS}_{\mathbb C[ x,\cdots]/\text{LT}(I)}~,
\ee
where ${\rm LT}(I)$ is the so-called \lq\lq leading ideal" of $I$. In the case of the MAD Schur ring, $R_{\infty}^{\rm MAD}$, $I$ is given in \eqref{arcspace}. The corresponding leading ideal is then
\be
\text{LT}(\langle{(x^2)^{(i)}\rangle}):=\langle{\left(x^{(i)}\right)^2~,\ x^{(i)}x^{(i+1)}\rangle}~.
\ee
In terms of the $\partial_+^iJ$ operators, this statement implies that \eqref{Jbasesapp} forms a basis for the space $V_{n,k}$.

We can check the consistency of this discussion with \eqref{HSinf} as follows. Define $B(n)$ to be the partition of $n$ into arbitrary parts differing by at least two. 
Then we have
\be
B(n)=\sum_{k=0}^\infty \dim V_{n-k,k}~,
\ee
where the shift in $n$ arises from the fact that we can have $n_1=0$ in \eqref{Jbasesapp} while $B(n)$ counts partitions with $n_1>0$. Finally, we can consider the corresponding partition function  (e.g., see (5) of \cite{alder1969partition})
\be\label{Bnpart}
\sum_{n=0}^\infty B(n)q^{n}
=\sum_{k=0}^\infty \frac{q^{k^2 } }{(q)_k}~,
\ee
which is consistent with \eqref{HSinf} after setting $p\to q$ \footnote{Note this partition function is the one appearing in the Rogers–Ramanujan identity.}.

\appendix*{\bf Supplemental Material III: Higher-rank theories}\label{higher}

In this section, we consider the $(A_1, A_{2r})$ SCFTs for general $r\ge1$ \cite{Cecotti:2010fi}. When $r=1$, we are back to the case of the MAD theory (i.e., ${\rm MAD}\cong(A_1, A_2)$).

These SCFTs are all quite similar to the $(A_1, A_2)$ theory. Their Macdonald index is \cite{Foda:2019guo}
\begin{eqnarray}\label{MDA1A2r}
 &&\mathcal{I}^{(A_1,A_{2r})}_M(q,T)= \sum_{n,k} \dim V_{n,k} q^{2k+n}T^k    
\cr&\!\!\!\! =\!\!\!\!\!\!&
\sum_{N_1\ge\cdots \ge N_r\ge 0}^\infty \frac{q^{N_1^2+\cdots N_r^2+N_1+\cdots+N_r}}{(q)_{N_1-N_2} \cdots (q)_{N_{r-1}-N_r} (q)_{N_r}} T^{N_1+\cdots +N_r}~. \!\!\!\!\!\!\nonumber\\  
 \end{eqnarray}
Explicitly for $r=1,2,3$, we have
\begin{eqnarray}
  \mathcal{I}^{(A_1,A_{2})}&=&
 1+T q^2+T q^3+T q^4+T q^5
   \cr && +\left(T+T^2\right) q^6+ \cdots, 
   \cr 
   \mathcal{I}^{(A_1,A_{4})}&=&1+T q^2+T q^3+\left(T+T^2\right) q^4+\left(T+T^2\right) q^5
   \cr && +\left(T+2 T^2\right) q^6+\cdots~,
    \cr 
 \mathcal{I}^{(A_1,A_{6})}&=&1+T q^2+T q^3+\left(T+T^2\right) q^4+\left(T+T^2\right) q^5
  \cr && +\left(T+2 T^2+T^3\right) q^6+\cdots~.  
\end{eqnarray}
 Moreover, the associated chiral algebras are $(2,2r+3)$ Virasoro minimal models \cite{Cordova:2015nma}
\begin{equation}\label{A1A2vir}
\chi((A_1, A_{2r}))={\rm Vir}_{c=-{2r(5+6r)\over3+2r}}~.
\end{equation}
Similarly to the  case of the MAD theory \eqref{MADRV}, the corresponding Zhu's algebra  is  now given by  
\be\label{RvA1A2r}
\mathcal R_{{\rm Vir}_{c=-{2r(5+6r)\over3+2r}}}=\mathbb C[x]/\langle x^{r+1}\rangle~.
\ee

Therefore, as described in the discussion section, since the corresponding hidden symmetry (Virasoro) is related to a conserved non-decoupling symmetry ($SU(2)_R$), it is natural to imagine that the bijection we saw between MAD Schur operators and free vector operators in (10) of the main text generalizes. Indeed, we will see this is the case.

Since the rank of the $(A_1, A_{2r})$ theory is $r$, the natural generalization of (4) of the main text is
\begin{equation}\label{TTmapSGEN}
\hat\CC_{0(0,0)}\ni J:=J^{11}_{+\dot+}\longrightarrow \Lambda_r:=\sum_{i=1}^r\lambda^1_{i,+}\bar\lambda^1_{i,\dot+}\in \hat\CC_{0(0,0)}^{\rm (Free)^{\times r}}~,
\end{equation}
and the natural generalization of our proposal in (10) of the main text is 
\begin{eqnarray}\label{proposalGen}
\CS_{(A_1,A_{2r})} &\ni& \partial^{i_1}_+J\cdots\partial_+^{i_n}J
 \cr&  \longrightarrow  &
   \partial^{i_1}_+\Lambda_r\cdots\partial^{i_n}_+\Lambda_r
 \cr& &\in\tilde\CS_{{\rm (Free\ Vector)}^{\times r}} 
  \subset\CS_{{\rm (Free\ Vector)}^{\times r}}~.\quad\qquad
\end{eqnarray}

It is straightforward to numerically check that \eqref{proposalGen} correctly reproduces the states computed by the Macdonald index. Since these are bosonic (as follows from the identification \eqref{A1A2vir}), this constitutes strong evidence of the proposal. Here we will give some analytic evidence as well.

Let us now argue for the map in \eqref{proposalGen}. Our first step is to understand the $(A_1, A_{2r})$ Schur ring. As in the case of (18) of the main text, the $(A_1, A_{2r})$ theory has a null vector involving $T_{2d}$ and its derivatives (except now at $h=2(r+1)$; this is the origin of \eqref{RvA1A2r}). The 4d interpretation of this null relation generalizes (19) of the main text
\begin{equation}
J^r(z) J(0)\supset J^{r+1}(0)=0~.
\end{equation}
This can also be understood from \eqref{RvA1A2r}. In the flow to the IR, this null relation follows from \eqref{TTmapSGEN} and Fermi statistics. 

As in the MAD case, it is natural to conjecture that the 4d Schur ring is generated by $\partial^i_+J$ subject to the constraint $J^{r+1}=0$. More precisely, the analog of \eqref{arcspace} is just
\begin{eqnarray}\label{arcspace2r}
X^{(A_1, A_{2r})}_\infty &=&\text{Spec } R^{(A_1, A_{2r})}_\infty~,\cr
R^{(A_1, A_{2r})}_\infty &=&\mathbb C[x^{(i)}  ]/ \langle{(x^{r+1})^{(i)} \rangle}~, \ \ \  i\ge 0~.
\end{eqnarray}

The idea is to compare the Hilbert series associated with \eqref{arcspace2r} with the Macdonald index. However, unlike the MAD case, the higher $r$ Hilbert series are not known in refined form. Instead, the unrefined case corresponding to $p=q$ is given by (see theorem 5.6 of \cite{bruschek2011arc})
\begin{eqnarray}\label{HSq2}
\text{HS} (q,p=q)\!\! &=&\!\!\sum_{n,k} \dim V_{n -k,k} q^n=\sum_{n,k} \dim V_{n  ,k} q^{n+k}
\cr &=&\prod_{i>0, \  i\neq 0 , r+1 ,  r+2 \!\!\!\mod \!(2r+3) }\frac{1}{1-q^i}~,\ \ 
\cr &=& 
\begin{cases}
1+q+q^2+q^3+2 q^4 +\cdots,&\!  r=1 \\
1+q+2 q^2+2 q^3+3 q^4+\cdots,& \! r=2 \\
1+q+2 q^2+3 q^3+4 q^4+\cdots,&\!   r=3\\
\qquad\vdots ~
\end{cases}\nonumber
\\
\end{eqnarray}

Therefore, we will need to compare this quantity with a somewhat unorthodox fugacity slice of the Macdonald index \eqref{MDA1A2r} gotten by setting $T=1/q$ (note that this is {\it not} the Schur index, where we would instead set $T=1$):
\begin{eqnarray}\label{qT1}
&&\mathcal{I}^{(A_1,A_{2r})}_M(q,T=1/q)= \sum_{n,k} \dim V_{n,k} q^{ n+k}\cr&=&\sum_{N_1\ge\cdots \ge N_r\ge 0}^\infty \frac{q^{N_1^2+\cdots N_r^2 }}{(q)_{N_1-N_2} \cdots (q)_{N_{r-1}-N_r} (q)_{N_r}}~.\qquad\quad
\end{eqnarray}
Intriguingly, these quantities can be written in terms of $r$-fold $q$-hypergeometric series \cite{zagier2007dilogarithm}. It would be interesting to understand if the refined index for $r>1$ can be expressed in terms of a generalization of \eqref{qhyper} (and to understand the connection between these various types of hypergeometric functions).

The equality of \eqref{HSq2} and \eqref{qT1} follows from the Andrews-Gordon identity. This is strong evidence that the 4d Schur ring is as described in \eqref{arcspace2r} (i.e., that it is generated by $\partial^i_+J$ subject to $J^{r+1}=0$).

We can then further conjecture the refined Hilbert series by setting $T\to p/q^2$ in the Macdonald index:
\be
\text{HS}(q,p)=\sum_{N_1\ge\cdots \ge N_r\ge 0}^\infty \frac{q^{N_1^2+\cdots N_r^2-N_1-\cdots-N_r}
p^{N_1+\cdots +N_r} }{(q)_{N_1-N_2} \cdots (q)_{N_{r-1}-N_r} (q)_{N_r}} ~.   \\
\ee
At low orders, we have verified this conjecture explicitly by enumerating the corresponding elements of the arc space \eqref{arcspace2r}.

To make contact with the IR description and the proposal in \eqref{proposalGen}, we again use leading ideals to generate a convenient UV basis for the arc space. To that end, we have (see proposition 5.2 of \cite{bruschek2011arc})
\begin{eqnarray}
\text{LT}(\langle{(x^{r+1})^{(i)}}\rangle)&=&\langle{ \big(x^{( j)}  \big)^s\big(x^{( j+1)}  \big)^{r+1-s}   \rangle}~,\cr&&\    j\ge 0~, \ \ \ s=0~,\ 1~,\ \cdots~,\ r~.\ \ \ \ \  \qquad
\end{eqnarray}
This discussion shows that a basis of operators  is given by
\begin{eqnarray}\label{Jbases2r}
&&(\p^{n_1} J)^{Q_1} \; (\p^{n_2 } J )^{Q_2} \;  \cdots ( \p^{n_s} J)^{Q_s}~, \ \ \  \cr&&
  \sum_{i=1}^s Q_i n_i=n~,  \ \ \qquad  \sum_{i=1}^s Q_i =k~,
\end{eqnarray}
where
\begin{eqnarray}\label{nQr}
&&0 \le n_1 < n_2 < \cdots  n_s~, \ \ \ Q_i \le r~, \cr&& Q_i+Q_{i+1}\le r \quad \text{ if }\quad  n_{i+1}=n_i+1~.
\end{eqnarray}
As a result, in \eqref{Jbases2r}, $n_i$ can repeat at most $r$ times. Moreover, $n_i$ and $n_i+1$ (if $n_i+1=n_{i+1}$) together can repeat at most $r$ times. From Gordon's Partition Theorem, the generating function of this partition is exactly given by \eqref{HSq2}.

To show that the map in \eqref{TTmapSGEN} does not affect the counting of the basis of states (and that therefore \eqref{arcspace2r} holds), we define
\be
P(i, m):=\p^{\lfloor{\frac{m }{2 }\rfloor}}\lambda^1 _{i,+} \p^{\lfloor{\frac{m+1 }{2 }\rfloor}}\bar\lambda^1 _{i,\dot+}
\in \p^m J~.
\ee
Then we have the following one-to-one correspondence
\begin{eqnarray}
&&\prod_{i=1}^s  (\p^{n_i} J  )^{Q_i}
\leftrightarrow
\prod_{i=1}^s
\Big[P(1, n_i) P(2, n_i)\cdots P(Q_i,n_i)\Big]^{ A_i}\cr&&
\times\Big[P(r, n_i) P(r-1, n_i)\cdots P(r-Q_i+1,n_i)\Big]^{ 1-A_i}~,\qquad \ \quad
\end{eqnarray}
where $A_i=i \mod 2$, and $A_{i+1}=1-A_i$. Note that the fermionic expression on the righthand side is never zero thanks to the condition \eqref{nQr}. Therefore, as in the MAD case, we have shown that we can also reproduce the Macdonald index from $r$ gauginos.
\end{appendix}

\bigskip

\end{document}